\documentclass[adp,fleqn]{w-art}
\usepackage{times}
\usepackage{w-thm}
\usepackage[]{graphicx}
\chardef\bslash=`\\ 

\begin{document}
\DOIsuffix{theDOIsuffix}
\Volume{12}
\Issue{1}
\Copyrightissue{01}
\Month{01}
\Year{2003}
\pagespan{1}{}

\keywords{Pentaquarks, solitons, diquarks, mixing, large $N_c$ limit.}
\subjclass[pacs]{11.30.Hv,12.39.Dc,14.20.Jn}



\title[Pentaquarks and SU(3)]{\flushright{\rm TPJU-15/2004}\\
\flushleft{SU(3) constraints on cryptoexotic pentaquarks}}
\author[M.\ Prasza{\l}owicz]{Micha{\l} Prasza{\l}owicz\footnote{
E-mail: {\sf michal@if.uj.edu.pl}, Phone: +48\,(12)\,632\,48 \,88,
Fax: +48\,(12)\,633\,40 \,79}\inst{1}}

\address[\inst{1}]{M. Smoluchowski Institute of Physics, Jagellonian
University, ul. Reymonta 4, Krak{\'o}w, Poland}



\begin{abstract}
We examine SU(3) constraints on the spectrum and decay widths of
cryptoexotic nucleon-like states in antidecuplet of pentaquarks.
We show that in the ideal mixing scenario the number of free
matrix elements describing splittings reduces from 4 to 2.
Model-independent sum rules are derived.
 Using as input $\Theta^+$ and $\Xi_{3/2}$ masses we show that
it is difficult to interpret Roper and $N^{\ast}(1710)$ nucleon
resonances as cryptoexotic pentaquarks. Large $N_c$ limit for
antidecuplet and accompanying octet in the diqaurk picture and the
analogy with the soliton model is also discussed.
\end{abstract}

\maketitle






\section{Introduction}

\label{sect1}

The discovery of the strange baryon $\Theta^+$ at 1540 MeV
\cite{Nakano:2003qx}\nocite{Barmin:2003vv,
Stepanyan:2003qr,Barth:2003es,
Asratyan:2003cb,Kubarovsky:2003fi,Airapetian:2003ri,
Aleev:2004sa,Abdel-Bary:2004ts,Aslanyan:2004gs}
- \cite{Chekanov:2004kn} opened a new chapter in hadron
spectroscopy. If confirmed by high statistics experiments, it will challenge
the commonly used naive quark model that worked so well for almost
four decades. Two questions will have to be answered: 1) what is the new
dynamical mechanism overlooked so far that renders light and narrow exotic
pentaquarks and 2) why the nonexotic states are not affected by this mechanism,
in other words: why the naive quark model works so well in the nonexotic sector.
Of course a likely answer might be that the key ingredient missing in the naive
quark model is chiral symmetry and that baryons are solitons in an effective
chiral  model \cite{Skyrme:1961vq}\nocite{Witten:1979kh} - \cite{Adkins:1983ya}.
After all, exotic antidecuplet  emerges naturally in chiral soliton models
\cite{Chemtob:1985ar} \nocite{Manohar:1984ys,Praszalowicz:1987em}
- \cite{Praszalowicz:2003ik}. Indeed,
light and narrow $\Theta^+$ was predicted many years
before its discovery \cite{Diakonov:1997mm} in the chiral quark-soliton model
\cite{Diakonov:1986yh}\nocite{Reinhardt:1989fv,Meissner:1989kq}
- \cite{Wakamatsu:1990ud}.
Even then some aspects of the soliton models have to be clarified and better
understood \cite{Itzhaki:2003nr}
\nocite{Cohen:2004xp,Cohen:2003yi,Pobylitsa:2003ju}
- \cite{Diakonov:2003ei}.

It is the purpose of the present paper to examine if the SU(3)
symmetry alone is able to explain, or -- more modestly -- to
accommodate further exotics that follow inevitably from the
discovery of $\Theta^+$. Experimentally another exotic state,
namely $\Xi_{3/2}^{--}(1860)$ (here subscript $3/2$ refers to
isospin),
 has been already announced by NA49 experiment at CERN \cite{Alt:2003vb}.
Theoretically, it is clear that there must be other cryptoexotic
states which span antidecuplet of flavor SU(3). These states are
cryptoexotic since unlike $\Theta^+$ and $\Xi_{3/2}$ their quantum
numbers can be constructed from 3 quarks only, however symmetry
requires that in fact they contain an additional $q \overline{q}$
pair. These cryptoexotic states, and more precisely the
cryptoexotic nucleon-like states, are the primary subject of the
present note. In Section \ref{sect2} we calculate mass spectrum in
the moderately broken SU(3) symmetry and then postulate ideal
mixing. Model-independent sum rules for the masses of exotic
baryons are derived. We also show that the ideal mixing ({\em
i.e.} requirement that physical states are defined as the ones
which posses definite number of (anti)strange quarks) is -- under
assumption of small symmetry breaking -- unable to accommodate
known nucleon resonances such as Roper and $N^{\ast}(1710)$. In
Sect. \ref{sect3} we further show that also the decay widths are
incompatible with ideal mixing.

In Sect.~\ref{sect4} we slightly change the subject. In the large
$N_c$ limit, flavor SU(3) representations are no longer ordinary
octet and decuplet. In soliton models their generalizations are
straightforwardly constructed
\cite{Diakonov:2003ei,Karl:1985qy,mp:largeN} because of the
constraint coming from the Wess-Zumino term
\cite{Guadagnini:1983uv}\nocite{Mazur:1984yf} -
\cite{Jain:1984gp}: $8=(1,1) \rightarrow (1,(N_c-1)/2)$ and $10
=(3,0) \rightarrow (3,(N_c-3)/2)$ \footnote{ Here we denote SU(3)
representations either by $(p,q)$ where number of quark indices is
$p$ and antiquark indices $q$, or by dimension with "bar" if
$q>p$.}. Moreover $\overline{10} =(0,3) \rightarrow
(0,(N_c+3)/2)$. We show that in the quark model with diquark
correlations, the construction in which diquarks contain 2 quarks
for arbitrary $N_c$ reproduces the same sequence of SU(3)
representations. As a result the ideal mixing can be generalized
to arbitrary (odd) $N_c$.

We summarize in Sect.~\ref{sect5}

Most of this work was done during my short visits at the Institut
f{\"u}r Theoretische Physik II of the Ruhr-University in Bochum in
January and June 2003. It is always a pleasure to visit this
lively international group led by Klaus Goeke whose support and
enthusiasm stay beyond the predictions which led to the discovery
of pentaquarks. It is a great honor to dedicate to Klaus this set
of remarks on the SU(3) nature of exotic states.

\section{SU(3) symmetry and ideal mixing}

\label{sect2}

Quantum numbers of $\Theta ^{+}$ require that its minimal quark content is $%
\left| uudd\overline{s}\right\rangle $. Two quarks can be either in flavor $%
\overline{3}$ or $6$. Therefore possible representations for 4 quarks are
contained in the direct products%
\begin{align}
\overline{3}\otimes \overline{3}& =3+\overline{6},  \nonumber \\
\overline{3}\otimes 6& =3+15,  \nonumber \\
6\otimes 6& =\overline{6}+15^{\prime }+15.  \label{twodiqs}
\end{align}%
Here $15=(2,1)$ and $15^{\prime }=(4,0)$.
Adding an $\overline{3}$ antiquark
yields%
\begin{align}
3\otimes \overline{3}& =1+8,  \nonumber \\
\overline{6}\otimes \overline{3}& =8+\overline{10},  \nonumber \\
15\otimes \overline{3}& =8+10+27,  \nonumber \\
15^{\prime }\otimes \overline{3}& =10+35,  \label{last}
\end{align}%
Therefore $\left| q^{4}\overline{q}\right\rangle $ state can be in one of
the following flavor representations:

\begin{equation}
\left| q^{4}\overline{q}\right\rangle \in1,\,8,\,10,\,\overline {10}%
,\,27,\,35.   \label{reps}
\end{equation}
Whether all representations (\ref{reps}) are allowed  depends
on the dynamics of a specific model.

\begin{vchfigure}[htb]
  \includegraphics[width=.5\textwidth]{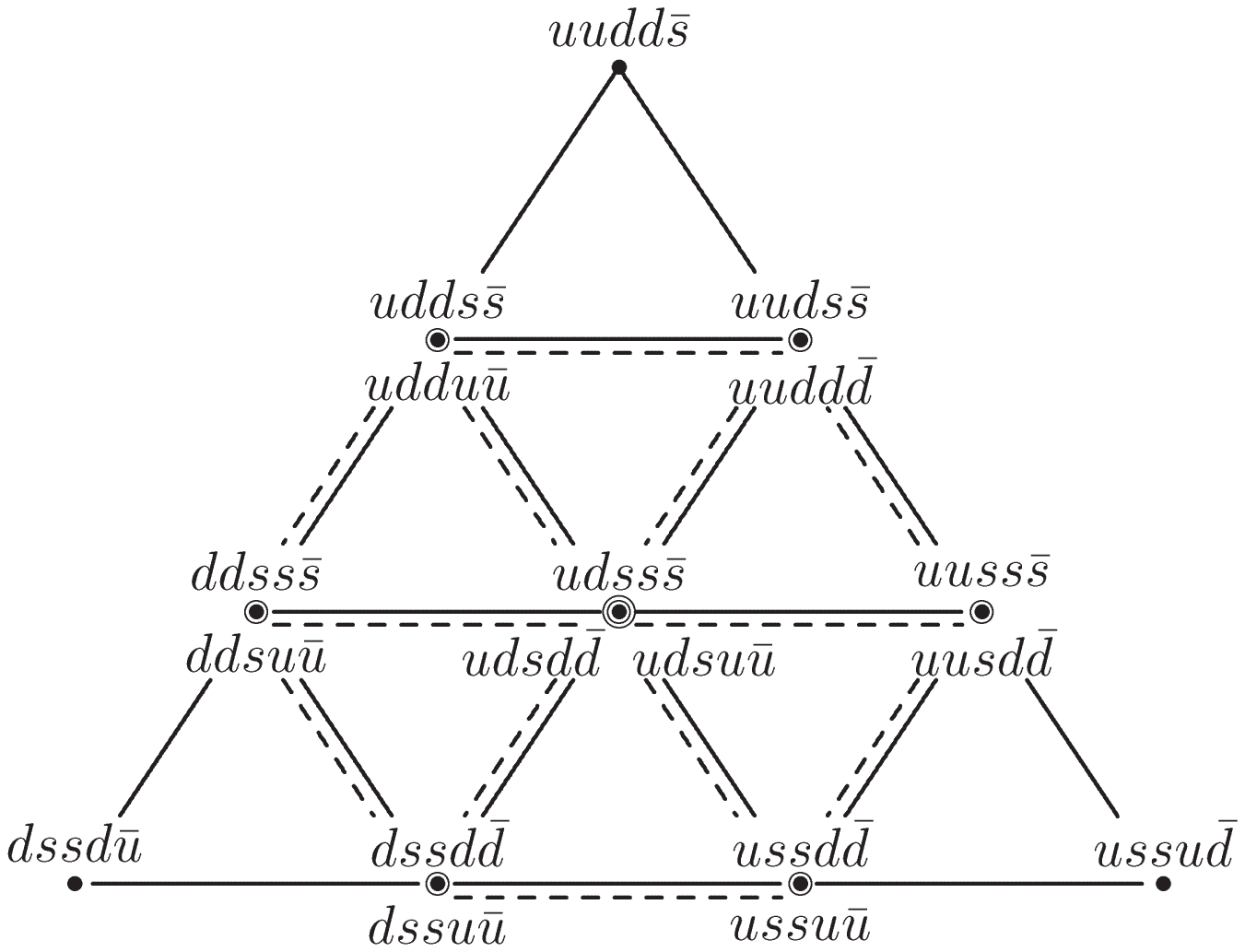}
\vchcaption{Mixing of $\overline{10}$ and $8$.}%
\label{fig:mixing}
\end{vchfigure}

Out of allowed representations (\ref{reps}) the lowest one including
explicitly exotic states is $\overline{10}$ which appears in a direct
product of 4 quarks in flavor $\overline{6}$ and an antiquark:
\begin{equation}
\bar{6}\otimes\bar{3}\rightarrow8+\overline{10}   \label{8and10bar}
\end{equation}
and is therefore inevitably accompanied by an octet (see Fig.~\ref%
{fig:mixing}). Since both representations may have spin $1/2$ it is
reasonable to assume that in the first approximation the masses of $8$ and $%
\overline{10}$ are degenerate. Unlike in the case of the ordinary octet and
decuplet, pentaquark symmetry states (\emph{i.e.} states which are pure
octet or antidecuplet) do not have a unique quark structure. For example a
proton-like state in antidecuplet and octet have the following quark content:%
\begin{align}
p_{\overline{10}} & =\sqrt{\frac{2}{3}}\left| uuds\overline{s}\right\rangle +%
\sqrt{\frac{1}{3}}\left| uudd\,\overline{d}\right\rangle ,  \nonumber \\
p_{8} & =\sqrt{\frac{1}{3}}\left| uuds\overline{s}\right\rangle -\sqrt{\frac{%
2}{3}}\left| uudd\,\overline{d}\right\rangle   \label{ideal}
\end{align}
where it is implicitly assumed that four quarks are in a pure $\overline{6}$
state. Similarly $\Sigma$-like states mix, while $\Theta^{+}$ and $\Xi$-like
states remain unmixed.

The QCD splitting hamiltonian $\Delta M$ that transforms like $Y=0,$ $I=0$
component of the octet tensor operator, has the following matrix elements
between the pentaquark states:%
\[
\left\langle 8,YII_{3}\left| \Delta M\right| 8,YII_{3}\right\rangle =D\left(
I(I+1)-\frac{1}{4}Y^{2}-1\right) -FY,
\]%
\begin{equation}
\left\langle \overline{10},YII_{3}\left| \Delta M\right| \overline {10}%
,YII_{3}\right\rangle =-CY   \label{diagonal}
\end{equation}
and%
\begin{equation}
\left\langle 8,YII_{3}\left| \Delta M\right| \overline{10}%
,YII_{3}\right\rangle =\left\langle \overline{10},YII_{3}\left| \Delta
M\right| 8,YII_{3}\right\rangle =\frac{G}{\sqrt{2}}   \label{8a10mix}
\end{equation}
for nucleon-like and $\Sigma$-like states. Constants $D,F,C$ and $G$ are
unknown reduced matrix elements. Equations (\ref{diagonal}) would be the
ordinary Gell-Mann--Okubo mass formulae if not for the mixing term (\ref%
{8a10mix}) which is non-zero because pentaquark octet and antidecuplet have
the same spin.

Now let us assume that strong interactions diagonalize (anti)strange quark
content. The off-diagonal elements of the mass matrix take the following
form for the nucleon-like states ($Y=1)$:%
\[
\left\langle udds\bar{s}\right| \Delta M\left| uddu\bar{u}\right\rangle
=\left\langle uuds\bar{s}\right| \Delta M\left| uudd\bar{d}\right\rangle =%
\frac{\sqrt{2}}{6}(-G-2C+D+2F)
\]
and for $Y=0,$ $I_{3}=\mp1$ $\Sigma$-like states:%
\[
\left\langle ddss\bar{s}\right| \Delta M\left| ddsu\bar{u}\right\rangle
=\left\langle uuss\bar{s}\right| \Delta M\left| uusd\bar{d}\right\rangle =%
\frac{1}{3\sqrt{2}}(G-2D)
\]
and similarly for $I_{3}=0$. The requirement that mass matrix is diagonal in
the (anti)strange quark basis leads to the relations:
\begin{equation}
G=2D,\quad C=F-\frac{D}{2}   \label{rel}
\end{equation}
so that the whole spectrum depends on two constants $D$ and $F$ and the
overall mass scale $M$:%
\begin{align}
\Theta^{+} & =M-2F+D,  \nonumber \\
N^{\ast} & =M-F-\frac{3}{2}D,  \nonumber \\
N_{s}^{\ast} & =M-F+\frac{3}{2}D,  \nonumber \\
\Sigma^{\ast} & =M-D,  \nonumber \\
\Sigma_{s}^{\ast} & =M+2D,  \nonumber \\
\Xi & =M+F-\frac{D}{2}.   \label{split}
\end{align}
Here by $N^{\ast}$ and $\Sigma^{\ast}$ we denote octet-like states with
additional $u\overline{u}$ or $d\overline{d}$ pair and subscript $s$ denotes
additional $s\overline{s}$ pair. $\Xi$ stands for both $I=1/2$ and $I=3/2$
states. The requirement that the mass matrix is diagonal in the
(anti)strange quark basis reduced the number of the unknown constants
describing the splittings from 4 to 2.

Scenario that the strong interactions diagonalize (anti)strange content
is known as \emph{ideal mixing}. It has been discussed in more detail in the
context of the diquark model proposed by Nussinov \cite{Nussinov:2003ex} and
Jaffe and Wilczek \cite{Jaffe:2003sg} where pentaquarks are viewed as bound
states of two diquarks and an antiquark. Diqaurks are color and flavor
antitriplets in a relative $P$-wave, hence two diquark state is a color $3$
and flavor $\overline{6}$. The motivation for a diqaurk picture comes from
the attraction between two quarks in a $\overline{3}$ color channel. Diquark
correlations have been discussed in the context of meson spectroscopy \cite%
{Jaffe:1977ig}. A schematic hamiltonian for a diquark-like
pentaquarks has been proposed in Ref.~\cite{Jaffe:2003sg}. It
contains two parameters: strange quark mass $m_{s}$ and parameter
$\alpha$ which is embodies the fact that diquarks involving
strange quark are more tightly bound. These
parameters are related to the reduced matrix elements introduced in (\ref%
{diagonal}) in the following way%
\begin{equation}
F=\frac{1}{6}(5\alpha+4m_{s}),\quad D=\frac{1}{3}(\alpha+2m_{s}).
\label{FDinJW}
\end{equation}
A somewhat different hamiltonian was introduced by Cohen in Ref.~\cite%
{Cohen:2004gu}. It is expressed in terms of a total number of strange quarks
and the excess of strange over antistrange quarks. the hamiltonian
$H=a(n_{s}+n_{\overline{s}%
})+b^{\prime}n_{s}$ also falls in the above category with $m_{s}\rightarrow a
$ and $\alpha\rightarrow b^{\prime}$ in (\ref{FDinJW}).

A quantitative support for the dipole picture comes also from the
instanton liquid model of the QCD vacuum. There, a tightly bound
scalar diquark is found to have a mass of a single constituent
quark, however, a tensor
diquark in flavor $6$ is only slightly heavier (570 MeV) \cite%
{Shuryak:2003zi}. If so, a bound state of scalar and tensor
diquarks and a light antiquark would fall into $8+10+27$ of flavor
SU(3) and, since there would be no penalty for the $P$-wave
excitation, this state should be lighter than a bound state of two
scalar diquarks in a $P$-wave and an antiquark. One should,
however, consider this possibility with care, since once the
diquarks overlap they may dissolve.

Any model hamiltonian of the SU(3) symmetry breaking with ideal mixing must obey
equations (\ref{split}). These equations lead to the model independent
relations%
\begin{align}
N_{s}^{\ast}-N^{\ast} & =\Sigma_{s}^{\ast}-\Sigma^{\ast},  \nonumber \\
2(N_{s}^{\ast}-\Theta^{+}) & =\Xi-N^{\ast},  \nonumber \\
N_{s}^{\ast}-\Sigma^{\ast} & =\Theta^{+}-N^{\ast}   \label{relations}
\end{align}
(and linear combinations of (\ref{relations})).

Now we shall try to constrain further the remaining two free
parameters $F$ and $D$. To this end we plot in
Fig.~\ref{fig:Dspectrum} the splittings (in units of $D$) as
functions of parameter $x=F/D$. We see that for $1/2<F/D<5/2 $
(depicted by thin vertical lines) the states are ordered according
to the increasing number of (anti)strange quarks. For $F/D=1$ the
levels are equally spaced and the states with the same number of
(anti)strange quarks are degenerate. It is therefore a reasonable
starting point for a small perturbation in $x$ which would lift
this degeneracy introducing small splittings in a way similar to
the $\Sigma - \Lambda$ splitting in the regular octet. We can see from Fig.~%
\ref{fig:Dspectrum} and Eqs.(\ref{split}) that to the left of $x=1$ the
following splittings become equal: $\Theta^+ - \Sigma^{\ast} =\Xi -
\Sigma^{\ast}$ for $x=7/10$, whereas to the right of $x=1$: $N_s -
\Sigma^{\ast} = \Xi - N_s$ for $x=3/2$. If the splittings of the states with
identical (anti)strange content are to be smaller than those which differ by
one (anti)strange quark, then one should require $7/10 < x < 3/2$.
We shall shortly see that
only $x>1$ leads to the reasonable phenomenology.

\begin{vchfigure}[htb]
  \includegraphics[width=.6\textwidth]{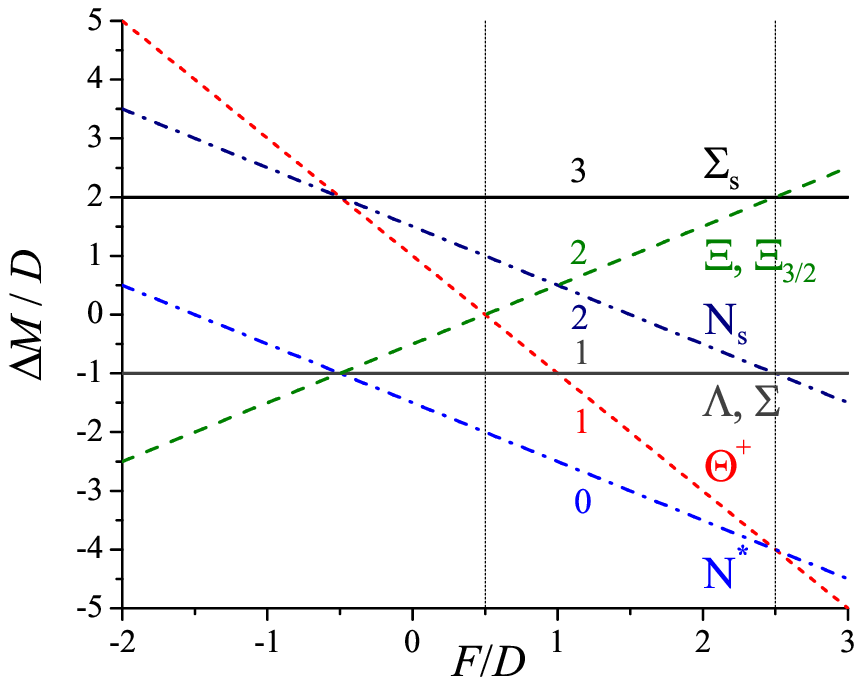}
\vchcaption{Pentaquark spectrum in the ideal mixing scenario as a
function of the parameter $x=F/D$. Numbers next to the lines
denoting different particles refer to the total number of strange
and antistrange quarks. For $1/2<x<5/2$ the states are ordered
according to the number of (anti)strange quark content.}
\label{fig:Dspectrum}
\end{vchfigure}

Now let us use experimental input for the masses of $\Theta^+$ and $\Xi_{3/2}
$. With this input the whole spectrum (\ref{split}) depends only on one free
parameter which we choose to be $x=F/D$. In Table (\ref{tab:xmass}) we
display the masses of the remaining cryptoexotic states and the values of
parameters $D$ and $F$ for three choices of $x$: 3/4, 1 and 3/2. We see that
for $x \le 1$ the lightest state, namely the $N^{\ast}$ resonance, is
unrealistically light. On the other hand for $x=3/2$ the two nucleon-like
resonances $N^{\ast}$ and $N_s^{\ast}$ have masses which allow to identify
them with the Roper and $N^{\ast}(1710)$.

\begin{align}
\begin{array}{c|ccc}
\hline
x=F/D & 3/4 & 1 & 3/2 \\ \hline
D\;[\text{MeV}] & 427 & 213 & 107 \\
F\;[\text{MeV}] & 320 & 213 & 161 \\ \hline
N^{\ast} & 793 & 1222 & 1433 \\
\Sigma,\,\Lambda & 1327 & 1540 & 1647 \\
N_{s}^{\ast} & 2074 & 1860 & 1754 \\
\Sigma_{s} & 2608 & 2180 & 1968 \\ \hline
\end{array}
\label{tab:xmass}
\end{align}

Although this seems encouraging, two remarks are here in order. First, $x=3/2
$ is really the upper bound and the spectrum is equally spaced with $\Delta
M \sim 107$ MeV. This contradicts our original assumption that perturbation
around $x=1$ should be small. The second remark concerns the decay widths
and will be discussed in the next Section.

\section{Decay widths in the ideal mixing scenario}

\label{sect3}

The decay width of a symmetry state $B$ in SU(3) flavor representation $R$
into an octet baryon $B^{\prime }$ and a pseudoscalar octet meson $\varphi $
reads:
\begin{equation}
\Gamma _{B\rightarrow B^{\prime }\varphi }=\frac{G_{R}^{2}}{8\pi
M\,M^{\prime }}\times C^{R}(B,B^{\prime },\varphi )\times p_{\varphi }^{3}.
\label{dw}
\end{equation}%
Here $M$, $M^{\prime }$ and $p_{\varphi }$ are baryon masses and meson
momentum in the decaying baryon reference frame. $C^{R}(B,B^{\prime
},\varphi )$ is the relevant SU(3) Clebsch-Gordan coefficient
(squared) and $G_{R}$
the decay constant. For states which are mixtures of different
representation states, like our candidates for Roper and $N^{\ast }(1710)$:
\begin{align}
\mathrm{Roper}& \rightarrow \left| N_{1}^{\ast }\right\rangle =\sqrt{\frac{1%
}{3}}\left| \overline{10},N\right\rangle -\sqrt{\frac{2}{3}}\left|
8,N\right\rangle ,  \nonumber \\
N^{\ast }(1710)& \rightarrow \left| N_{2}^{\ast }\right\rangle =\sqrt{\frac{2%
}{3}}\left| \overline{10},N\right\rangle +\sqrt{\frac{1}{3}}\left|
8,N\right\rangle ,  \label{Roper1710}
\end{align}%
the amplitude for the decay width would be the sum over $R=\overline{10}$
and 8. However, since the decay width of $\Theta ^{+}=\left| \overline{10}%
,\Theta ^{+}\right\rangle $ is very small, indicating that $G_{\overline{10}%
}\sim 0$ (remember $\Theta^+$ does not mix),
the $\overline{10}$ component in Eqs.(\ref{Roper1710}) can be
safely neglected in the first approximation. Then
\begin{equation}
\frac{\Gamma _{N_{1}^{\ast }\rightarrow N\pi }}{\Gamma _{N_{2}^{\ast
}\rightarrow N\pi }}\sim 2\frac{M_{2}}{M_{1}}\frac{p_{1}^{3}}{p_{2}^{3}}\sim
0.75  \label{Gamrat}
\end{equation}%
where we have used physical masses for Roper and $N^{\ast
}(1710)$. Equation (\ref{Gamrat}) indicates that partial decay
widths of the nucleon-like pentaquark resonances have to be of the
same order: either both small or both large. This observation was
first made by Cohen \cite{Cohen:2004gu} and presented in form of
the inequality connecting different decay constants.
Experimentally \cite{Cohen:2004gu} $\Gamma
_{\text{Roper}\rightarrow N\pi }\sim 228$ MeV and $\Gamma
_{1710\rightarrow N\pi }\sim 15$ MeV. Order of magnitude
difference between the partial decay widths of Roper and $N^{\ast
}(1710)$ remains in contradiction with the ideal mixing scenario.

Of course ideal is mixing probably an idealization.
The nonideal mixing would invalidate Cohen's inequality.
More general scenarios
have been discussed in Refs.~\cite{Diakonov:2003jj}
\nocite{Pakvasa:2004pg,Arndt:2003ga,Ellis:2004uz} -
\cite{Praszalowicz:2004dn}.
However, the discussion of masses and
decay widths of the $N^{\ast }$ states under assumption that they
correspond to the Roper and $N^{\ast }(1710)$ done in Ref.~\cite%
{Pakvasa:2004pg} still indicates that it is impossible to match the mass
splittings with the observed branching ratios for these two
resonances even for arbitrary mixing. It is shown that the
\emph{nonideal}\ mixing required for the decay $N^{\ast
}(1710)\rightarrow \Delta \pi $ is not compatible
with the mixing deduced from the masses. The conclusion of Ref.~\cite%
{Pakvasa:2004pg} is that most probably $\Xi _{3/2}(1860)$ is not a member of
$\overline{10}$ but rather of $27$. That would release constraints on the $%
N_{\overline{10}}$ state coming from the equal spacing of
antidecuplet. Another possibility based on the nonideal mixing
scenario advocated by
Diakonov and Petrov \cite{Diakonov:2003jj} is that there should be a new $%
N^{\ast }$ resonance in the mass range of $1650\div 1680$ MeV, a
possibility discarded in Ref.~\cite{Pakvasa:2004pg}.

\section{Large $N_c$ limit: solitons vs. diquarks}

\label{sect4}

\begin{vchfigure}[htb]
\vspace{0.1cm}
  \includegraphics[width=0.8\textwidth]{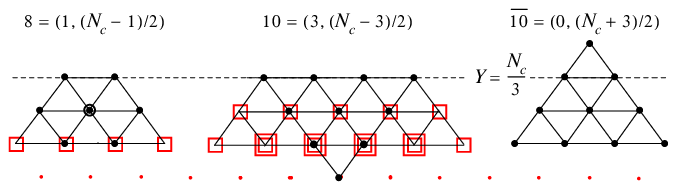}
\vchcaption{SU(3)-flavor representations for arbitrary $N_{c}$. Representations
are selected by a demand that they contain states with $Y=N_{c}/3$. Spurious
states which disappear for $N_{c}=3$ are denoted by open squares.}%
\label{fig:largereps}
\end{vchfigure}

For arbitrary $N_{c}$ ordinary baryons consist of $N_{c}$ quarks
and ''pentaquarks'' are built from $N_{c}+1$ quarks and an
antiquark. In the soliton picture relevant flavor representations
are selected by the requirement that physical multiplets
contain states with hypercharge $Y=N_c/3$, leading to the
following generalizations
\cite{Diakonov:2003ei,Karl:1985qy,mp:largeN}:
\begin{align}
8=(1,1) & \rightarrow (1,\frac{N_c-1}{2}), \nonumber \\
10 =(3,0) & \rightarrow (3, \frac{N_c-3}{2}), \nonumber \\
\overline{10} =(0,3) & \rightarrow (0, \frac{N_c+3}{2})
\label{eq:largereps}
\end{align}
as illustrated in Fig.~\ref{fig:largereps}.

Although it possible to show on general grounds that
representation content of the quark model and soliton model
coincide for large $N_c$ \cite{Manohar:1984ys,Jenkins:2004tm}, it
is instructive to illustrate this in the specific quark model.
Moreover, the general arguments apply to the highest
representations, {\em i.e.} antidecuplet in the case of
"pentaquarks" but not to the accompanying octet. In the following
we shall construct large $N_c$ generalizations of the flavor
representations for "pentaquarks" in the diquark picture and show
that they coincide with (\ref{eq:largereps}).

For "pentaquarks" in the
diquark picture we would have $(N_{c}+1)/2$ diquarks (remember $N_{c}$ is
odd). For $N_{c}>3$ two quarks antisymmetrized in color and flavor still
form flavor $\overline{3}$ but color representation is (anti) $%
N_{c}(N_{c}-1)/2$. Here we denote SU(3) representations by the number of
quark and antiquark indices $(p,q)$ and/or for SU($N_{c}$) by dimension with
a suffix anti- (or bar) if $q>p$. Generalizing scenario of Jaffe and Wilczek %
\cite{Jaffe:2003sg} we put all diquarks in a color and space
antisymmetric state. We can easily antisymmetrize $N_{c}-1$ quarks
leaving the last diquark aside. Adding this last diquark
(antisymmetrized quark pair) results in two color structures, one
of them being $N_c$. This is depicted in
Fig.\ref{fig:colorN}. So adding an antiquark in color
$\overline{N}_{c}$ will result in a
color singlet state.

\begin{vchfigure}[htb]
  \includegraphics[scale=0.6]{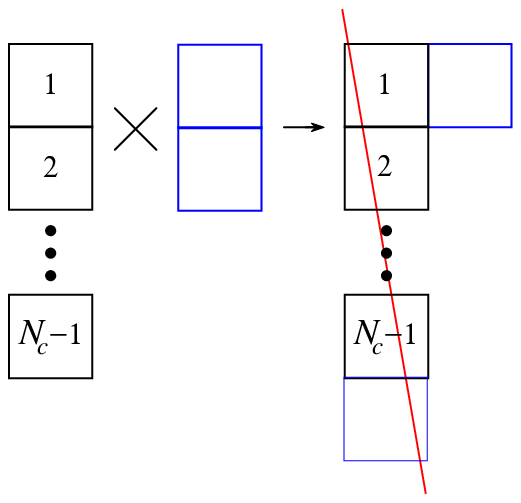}
\vchcaption{Color representation for $(N_{c}+1)/2$ diquarks. First, $N_{c}-1$
quarks are symmetrized and then the last two quarks are added in a form of a
diquark to produce fundamental representation of SU($N_{c}$).}%
\label{fig:colorN}
\end{vchfigure}

In flavor space we have to symmetrize all diquarks forming
representation $(0,(N_c+1)/2)$. Adding an antiquark in flavor
$\overline{3}=(0,1)$, as depicted in Fig.~\ref{fig:flavorN},
results in two representations: $"8"=(1,(N_c-1)/2)$ and
$"\overline{10}"=(0,(N_{c}+3)/2)$. These are the same
representations as in the soliton case (\ref{eq:largereps}).

\begin{vchfigure}[htb]
  \includegraphics[scale=0.6]{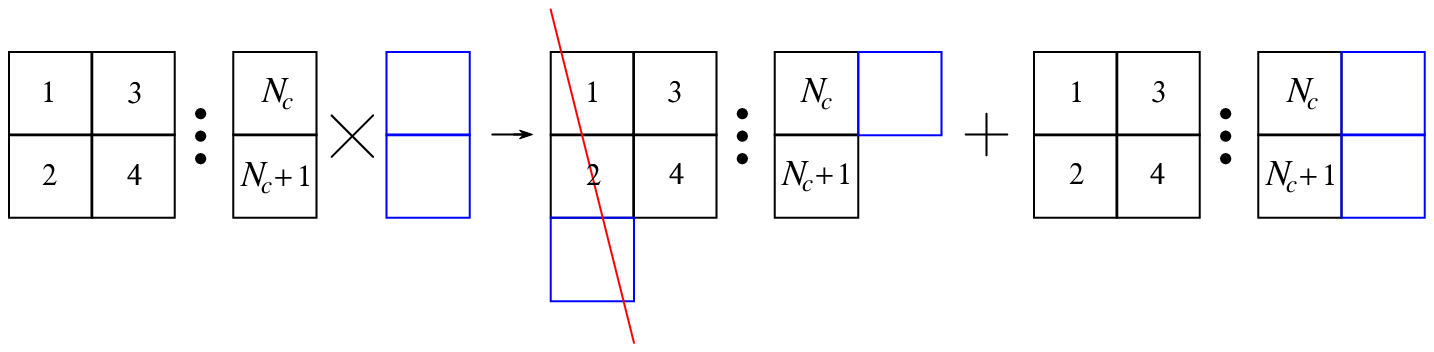}
\vchcaption{Flavor representations for $(N_{c}+1)/2$ diquarks and an antiquark.
Diquarks are symmetrized and adding an antiqauark produces $"8"=(1,(N_{c}%
-1)/2) $ and $"\overline{10}" =(0,(N_{c}+3)/2)$.}%
\label{fig:flavorN}%
\end{vchfigure}

Note that this generalization of octet and antidecuplet allows for
ideal mixing scenario. Indeed, both nucleon-like and $\Sigma$-like
states mix in the same way as for $N_c=3$. Only $\Xi_{3/2}$ in
$"\overline{10}"$ will additionally mix with spurious sates in
$"8"$, however, the latter states will disappear for $N_c=3$.

Finally let us remark, that another generalization of the diquark
picture which parallels junction models would result in a
completely different representation structure. Indeed, another
possible generalization of a diquark would be to antisymmetrize
$N_c-1$ quarks. We would have then to take $N_c-1$ of such
"diquarks" and add an antiquark. In this case "pentaquarks" would
consist of $(N_c-1)^2$ quarks and an antiquark. This would be a
completely different object than the regular pentaquark
constructed from $N_c+1$ quarks and an antiquark, although
accidentally for $N_c=3$ both pictures coincide.

\section{Summary}

\label{sect5}

Present study shows that consistent description of cryptoexotic nucleon-like
states requires new $%
N^{\ast }$ resonances in the mass range of $1650\div 1680$ MeV.
Similar conclusion has been reached in Ref.~\cite{Arndt:2003ga} where mixing
scenario was discussed within the framework of the quark soliton model.
Here already the ordinary nucleon state has a non-negligible admixture of $%
\overline{10}$. Taking this mixing into account the authors of Ref.~\cite%
{Arndt:2003ga} estimated the width of a possible but yet undiscovered $%
N^{\ast }$ state of a mass of $1680$ $(1730)$ MeV to be $\Gamma
_{N^{\ast }\rightarrow N\pi }\sim 2.1$ $(2.3)$ MeV. Although quite
narrow, this width is too large to be accommodated in the $\pi N$
scattering data. The same authors \cite{Arndt:2003ga} claim that
the improved phase shift analysis admits two candidates for the
narrow resonances of these masses but with decays widths smaller
than $0.5$ ($0.3$) MeV. Further decrease of theoretical
predictions might be achieved by adding a mixing to yet another
nucleon-like state as Roper and $N^{\ast }(1710)$. And even
further suppression of this decay is provided by the $27$
admixture as discussed in
Ref.~\cite{Ellis:2004uz,Praszalowicz:2004dn}.

We have also shown that a particular construction of SU(3) flavor
baryon representations
that generalizes the diquark model for large $N_c$, coincides
with the soliton model.

To conclude  let us note that physics of $N^{\ast }$ and $\Sigma
^{\ast }$ states will be most probably physics of extensive mixing between
different nearby states. It is clear from our discussion above that the
consistent physical picture requires the existence of yet undiscovered
nucleon resonances within the mass range of $1650-1750$ MeV or so.
Experimental searches for new narrow nucleon-like states have been recently
performed by some experimental groups with positive preliminary evidence %
\cite{Kabana:2004trans,Kouznetsov:2004trans}

\begin{acknowledgement}
The author acknowledges discussions with T. Cohen, D.I. Diakonov, K. Goeke, R.L. Jaffe,
V.Yu. Petrov, P.V. Pobylitsa, and M.V. Polyakov.
The present work is supported
by the Polish State Committee for Scientific Research (KBN) under grant 2 P03B
043 24.
\end{acknowledgement}


\end{document}